%% file: nf.tex
\documentclass{JHEP3}

\usepackage{epsfig,multicol}

\newcommand{\be}{\begin{equation}}
\newcommand{\ee}{\end{equation}}
\newcommand{\bea}{\begin{eqnarray}}
\newcommand{\eea}{\end{eqnarray}}
\newcommand{\beann}{\begin{eqnarray*}}
\newcommand{\eeann}{\end{eqnarray*}}
\newcommand{\nn}{\nonumber}
\newcommand{\ba}{\begin{array}}
\newcommand{\ea}{\end{array}}

\newcommand{\Tr}{{\rm Tr}}
\newcommand{\diag}{{\rm diag}}
\newcommand{\Vol}{{\rm Vol}}
\newcommand{\Qt}{\tilde{Q}}
\newcommand{\tv}{\mbox{\boldmath$t$}}
\newcommand{\vv}{\mbox{\boldmath$v$}}
\newcommand{\av}{\mbox{\boldmath$a$}}
\newcommand{\del}{\partial}

\title{Exact Mesonic Vacua From Matrix Models}

\author{Kazutoshi Ohta\\
Department of Physics, University of Wales Swansea,
Swansea, SA2 8PP, U.K.\\
E-mail: \email{k.ohta@swansea.ac.uk}
}

\preprint{\hepth{0212025}, SWAT-360}

\abstract{
We investigate in detail the structure of mesonic vacua of $N{=}1$ $U(N_c)$ supersymmetric gauge theory with $N_f$ flavors from the matrix model. We show that the Witten index from the matrix model calculation agrees with a result from field theoretical analysis. We also discuss the relationship between a diagrammatic summation and direct matrix integration with insertion of a variable changing delta function. Using this formalism, we obtain the quantum moduli space and evidence of the Seiberg duality from the matrix models.
}


\begin{document} 

\section{Introduction}

Dijkgraaf and Vafa recently showed that an effective superpotential of $N{=}1$ supersymmetric gauge theory is obtained from a hermitian matrix integral over a tree level superpotential \cite{DV1,DV2,DV3}. This observation is originally formulated by using a string theoretical realization of gauge theory, but even apart from string theory the Dijkgraaf-Vafa (DV) procedure itself produces field theoretically a powerful principle to calculate the effective glueball superpotential. After their works, a lot of papers, which discuss from various point of view, have rapidly appeared in \cite{DHKS1}--\cite{ACHKR}.

The model discussed in \cite{DV1,DV2,DV3} is mainly the $N{=}1$ gauge theory coupled to adjoint chiral superfields with a general superpotential of a polynomial. And also $N{=}1^*$ and Leigh-Strassler deformed theory from $N{=}4$ theory are discussed in detail by \cite{DHKS1,DHKS2,DHK}. On the other hand, theory with chiral fields in a fundamental representation, which is the so-called supersymmetric QCD (SQCD) theory, has much interesting dynamics like a generation of the Affleck-Dine-Seiberg (ADS) superpotential \cite{ADS}, Seiberg duality \cite{Seiberg1,Seiberg2,IS}, etc. So we expect that the matrix model technique sheds new light on the non-perturbative dynamics of $N{=}1$ SQCD and phenomenologically realistic models.

A breakthrough in the calculation of matrix models with flavors has been achieved by \cite{ACFH}. They find that for $U(N_c)$ gauge theory with $N_f$ flavors in the fundamental representation, the matrix integral is completely performed and infinite series in a perturbative expansion could be represented by a single analytic function. Using their results, a generation of ADS superpotential is discussed in \cite{Suzuki,BR}. However, these matrix model calculation has a discrepancy in a sense of the number of supersymmetric vacua (Witten index) compared with results of \cite{CKM} except for the case of $N_f=1$ \cite{Suzuki}. This implies that we overlook some information in the matrix model calculation.

Moreover, the generation of the ADS superpotential is discussed in a different way \cite{DJ} by insertion of a matrix valued delta function, which is called as the Wishart integral in the mathematical literature \cite{Fyodorov,JN}. However, the relationship between these two different methods is still unclear. If we naively use the result from the Wishart integral for the $N_c < N_f$ case, the matrix integral indicates an existence of the Seiberg duality \cite{Feng,FH}. But we need more careful analysis for this attractive problem.

The purpose of this paper is to resolve the discrepancy of the Witten index between the effective mesonic field theory and matrix model calculation, and to find an exact correspondence between the perturbative series expansion and the direct matrix integration with insertion of the delta function. The former is just done by perceiving ambiguity of the functional representation in a perturbative infinite series. The latter can be found by refining the arguments in \cite{DJ} from the first principle of the matrix integral. In our argument, we obtain the exact ADS superpotential without the shift of the mass term for the meson field.

The paper is organized as follows. In the subsequent section, we extend the analysis of the effective mesonic vacua which is discussed in \cite{CKM} for the $SU(N_c)$ case. This analysis tells us a disagreement of the Witten index with the previous matrix model calculation. We reproduce the matrix integration for a general mass matrix in section 3 and we show one to one correspondence between the vacua of the meson and glueball effective theory. In section 4, we discuss in detail the relationship between the results of the perturbative sum of the matrix integration and one from the direct matrix integration. We will give a meaning of the mesonic effective theory in terms of the matrix models.

\section{Vacuum structure of effective meson theory}

We consider $U(N_c)$ gauge theory with $N_f$ flavors in the fundamental representation in the following. The tree level superpotential of this theory is obtained by a flow from $N{=}2$ SQCD theory with fundamental flavors $Q_i^a$ and $\Qt_i^a$ by adding a mass $\mu$ for the adjoint scalar $\Phi_a^b$ in $N{=}2$ vector multiplet
\be
W_{\rm tree}(\Phi,Q,\Qt)=\mu \Tr \Phi^2
+ \sqrt{2}\Qt^a_i \Phi_a^b Q^i_b
+ m_i \Qt_i^a Q_a^i,
\label{Wtree}
\ee
where
$i,j=1,\ldots,N_f$ stand for the flavor indices and $a,b=1,\ldots,N_c$ stand for the color indices.


If we integrate out $\Phi$ first, we expect the effective superpotential for mesons from the principle of holomorphy and dimensional counting as follows
\be
W_{\rm eff}(M) = -\frac{1}{2\mu}
\Tr M^2
+\Tr(mM)
+W_{\rm ADS},
\label{eff M}
\ee
where we define the gauge singlet meson variable by $M_i^j\equiv Q^a_i \Qt_a^j$ and $W_{\rm ADS}$ is an instanton induced effective ADS superpotential proposed by \cite{ADS}
\be
W_{\rm ADS} = (N_c-N_f)
\left(
\frac{\Lambda_1^{3N_c-N_f}}{\det M}
\right)^{\frac{1}{N_c-N_f}},
\label{ADS}
\ee
with the scale of the $N{=}1$ SQCD theory $\Lambda_1$. The vacuum structure of this effective theory is carefully analyzed in \cite{CKM} for $SU(N_c)$ theory. We follow their arguments below for the $U(N_c)$ gauge group.

Using the $U(N_f)$ global symmetry, we can diagonalize $M$ as
\be
M = \diag(\lambda_1,\lambda_2,\ldots,\lambda_{N_f}).
\label{diagM}
\ee
Substituting the diagonalized $M$ into the effective superpotential (\ref{eff M}), we have the vacuum equations for each $\lambda_i$
\be
-\frac{1}{\mu}
\lambda_i
+m_i
-\frac{\Lambda_1^{(3N_c-N_f)/(N_c-N_f)}}{(\prod_j \lambda_j)^{1/(N_c-N_f)}}
\lambda_i^{-1}=0,
\ee
and equivalently this reduces to a quadratic equations
\be
\lambda_i^2
-\mu\, m_i\lambda_i
+\mu X=0,\label{quadra1}
\ee
where we define
\be
X \equiv \Lambda_1^{(3N_c-N_f)/(N_c-N_f)}
(\prod_j \lambda_j)^{1/(N_f-N_c)}.
\label{def X}
\ee

These quadratic equations have two different solutions for each $\lambda_i$. We now introduce two indices $i$ and $i'$ which run for solutions with negative and positive signs, respectively, within $N_f$ indices (this choice is for later convenience), that is, we choose
\bea
\lambda_i
 &=& \frac{\mu\,m_i}{2}\left(1 - \sqrt{1-4\alpha_i X}\right),
\quad \mbox{$i$ runs for $r$ within $N_f$}
\label{lambdais1}\\
\lambda_{i'}
 &=& \frac{\mu\,m_{i'}}{2}\left(1 + \sqrt{1-4\alpha_{i'} X}\right),
\quad \mbox{$i'$ runs for $N_f{-}r$ within $N_f$},
\label{lambdais2}
\eea
where $r$ is the number of the solutions with negative signature and we
define $\alpha_i\equiv \frac{1}{\mu m_i^2}$. 

We now recall that $X$ is defined in (\ref{def X}) in terms of $\lambda_i$'s. So putting (\ref{lambdais1}) and (\ref{lambdais2}) into the definition of $X$, we obtain the following equation for each fixed $r$
\be
X
=\Lambda_1^{\frac{3N_c-N_f}{N_c-N_f}}
\left(\det\left(\frac{\mu\,m}{2}\right)\right)^\frac{1}{N_f-N_c}
\prod_i \left(1-\sqrt{1-4\alpha_i X}\right)^\frac{1}{N_f-N_c}
\prod_{i'} \left(1+\sqrt{1-4\alpha_{i'} X}\right)^\frac{1}{N_f-N_c},
\ee
and reduce to
\be
X^{N_c}
=\frac{1}{2^{N_f}}\Lambda^{3N_c}
\prod_i \left(1+\sqrt{1-4\alpha_i X}\right)
\prod_{i'} \left(1-\sqrt{1-4\alpha_i X}\right),
\label{X=prod}
\ee
where $\Lambda^{3N_c}\equiv (\det\,m) \Lambda_1^{3N_c-N_f}$ is a dynamical scale of the pure $U(N_c)$ gauge theory.

We consider in the following the case that the mass matrix is proportional to an identity in order to see the explicit structure of solutions in eq.~(\ref{X=prod}). In this simple case, the general solution of the eigenvalues preserves a global $U(r)\times U(N_f-r)$ symmetry and eq.~(\ref{X=prod}) becomes
\be
X^{N_c} = \frac{1}{2^{N_f}}\Lambda^{3N_c}\left(1+\sqrt{1-4\alpha X}\right)^r
\left(1-\sqrt{1-4\alpha X}\right)^{N_f-r},
\label{identity case}
\ee
where
all $\alpha_i$'s are identical to $\alpha$ now.
If we assume $r>N_f{-}r$ without loss of generality, we can rewrite the vacuum equation into
\be
Y^{N_c-r}(1-Y)^{N_c-N_f+r}=\Lambda^{3N_c}\alpha^{N_c},
\label{reduced vacuum eq}
\ee
where we define $Y \equiv (1+\sqrt{1-4\alpha X})/2$.


We can see that eq.~(\ref{reduced vacuum eq}) is a polynomial of degree $2N_c{-}N_f$ in $Y$. So we expect that eq.~(\ref{identity case}) has $2 N_c{-}N_f$ solutions for each fixed $r$ as well as eq.~(\ref{reduced vacuum eq}). The number of ways for choosing $r$ negative signature is given by ${}_{N_f}C_r$, but if we note that the polynomial (\ref{reduced vacuum eq}) is invariant under exchanging the signs of $r$ positive roots with $N_f{-}r$ negative ones in eq.~(\ref{identity case})\footnote{There is no symmetry like this for the general mass matrix.}, the essential number of choice is a half of these. So we can count, in terms of $X$, the total number of vacua (Witten index) as
\bea
\Tr (-1)^F &=& \frac{1}{2}(2N_c-N_f)\sum_{r=0}^{N_f}{}_{N_f}C_r\nn\\
&=& (2N_c-N_f) 2^{N_f-1}.
\label{WittenM}
\eea


In comparison with the results in \cite{ACFH,BR}, there exists a discrepancy of the Witten index with the factor of $2^{N_f-1}$ except for the $N_f{=}1$ case \cite{Suzuki}. We will refine the matrix model calculation in the next section to resolve this discrepancy.

%
%
%
%

\section{Matrix integral over flavors}

We will discuss the derivation of the glueball effective theory from the matrix model procedure given by \cite{DV1,DV2,DV3}. We here extend the diagrammatic calculation with flavors in \cite{ACFH} to a general mass matrix.

As conjectured in \cite{ACFH}, only the diagram with one flavor index loop contributes to the effective superpotential. A topology of this diagram is a disk with one boundary and we have to sum up all other color index holes. We can insert $S=\Tr \lambda^2$ into each color index holes except for one index loop in order to cancel fermionic zero modes. In the diagram which contains only the color index loop, we can choose one hole in which no $S$ operator is inserted among total $h$ holes. Therefore, the contribution to the superpotential is proportional to $N_c h S^{h-1}$, see Fig.~1, where $h$ in the prefactor is the number of choices of no operator inserted loop and the factor $N_c$ comes from the trace of Chan-Paton color index. This means that the contribution to the effective superpotential from the free energy of the matrix model is $N_c\frac{\del {\cal F}_{\chi=2}(S)}{\del S}$, where ${\cal F}_{\chi=2}(S)$ is a sum of the planar diagram with the Euler number $\chi{=}2$, or the genus zero.

On the other hand, if we consider the diagrams with one flavor index loop (boundary), we can not insert the operator $S$ into the flavor loop. So we must insert into $h-1$ color indexed holes and there is no other choice of the insertion. This gives just the contribution of $S^{h'}$ to the superpotential, where $h'=h-1$. So the contribution to the superpotential directly comes from a sum of the diagram with one boundary ${\cal F}_{\chi=1}(S)$ instead of the derivative of the free energy. We finally take a trace over the flavor index. This leads to the factor $N_f$ if the system possesses a $U(N_f)$ global symmetry, but it becomes a sum over the different function with each flavor indices in general, namely $\sum_{i=1}^{N_f} {\cal F}_{\chi=1}^{(i)}(S)$, where each function depends on some parameters, which break the global symmetry, like different masses.

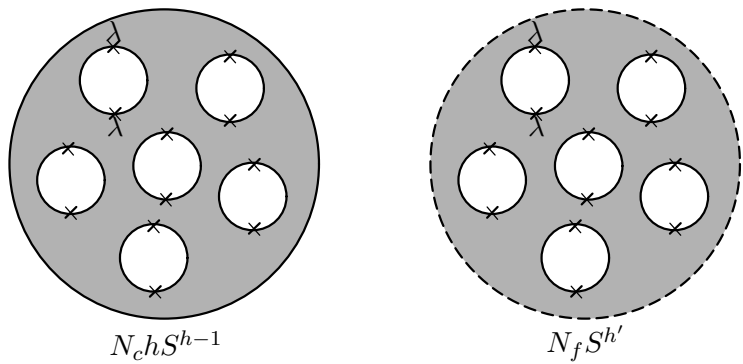
\begin{figure}[t]
  \begin{center}
    \input{fig.tex}
  \end{center}
  \caption{Open string world sheet diagrams of the flavor theory that contribute to the superpotential. We denote color index loops as solid lines and a flavor index loop as a dotted line. There is no operator $S$ insertion in the outer index loop. The contribution from the diagram with one flavor index loop (one boundary) is proportional to $N_f S^{h'}$.}
\end{figure}

As mentioned above, if we use a mass matrix which is proportional to an identity matrix and preserves an $U(N_f)$ global symmetry, we have only the contribution from the diagrams with one boundary is always proportional to $N_f$. This hides the detailed vacuum structure from the matrix model calculation. So we extend the calculation of \cite{ACFH} to the general mass matrix, which is diagonalized by the global symmetry at least, to compare with the field theoretical results in the previous section.

Now let us consider the matrix integral with the tree level superpotential (\ref{Wtree}) to evaluate the contribution from the diagrams with one boundary,
\be
Z=\frac{1}{\Vol(U(N_c))}\int [d\Phi][dQ][d\Qt]
\exp\left\{
-\frac{N_c}{S}W_{\rm tree}(\Phi,Q,\Qt)
\right\},
\label{mat int}
\ee
where $\Vol(U(N_c))$ is a volume of the gauge group $U(N_c)$. We regard the quadratic terms of the matrices in $W_{\rm tree}(\Phi,Q,\Qt)$ as a kinetic term and the Yukawa interaction term as an interaction of the perturbative expansion. We find two point correlators (propagators) from the tree level superpotential as
\be
\quad
\langle \Phi_a^b \Phi_c^d \rangle
= \frac{1}{2\mu}\frac{S}{N_c} \delta_a^d \delta^b_c,
\quad
\langle Q^a_i \Qt_b^j \rangle
=\frac{1}{m_i}\frac{S}{N_c}\delta^a_b \delta_i^j.
\ee

The contribution to ${\cal F}_{\chi = 2}(S)$ comes only from the volume factor and the measure of a gaussian integral of $\Phi$ since $W_{\rm tree}$ does not include any self-interacting part of $\Phi$. So the matrix integral (\ref{mat int}) has the following series expansion with ignoring the volume factor
\be
Z=\sum_{k=0}^{\infty}\frac{1}{(2k)!}
\left(\frac{\sqrt{2}N_c}{S}\right)^{2k}
\langle
(\Qt\Phi Q)_1(\Qt\Phi Q)_2\cdots(\Qt\Phi Q)_{2k}
\rangle.
\ee
Using the Feynman rules obtained from the superpotential, we can find weights of the connected diagrams with one boundary and contract each two point correlators, thus we get
\be
\frac{N_c}{S}{\cal F}_{\chi=1}(S,\mu,m_i)
=-\sum_{k=1}^{\infty}\frac{(2k-1)!}{(k+1)!k!}
\left(\frac{\sqrt{2}N_c}{S}\right)^{2k}
\left(\sum_{i=1}^{N_f}\left(
\frac{1}{m_i}\frac{S}{N_c}\right)^{2k}
\right)
\left(\frac{S}{2\mu N_c}\right)^{k}
N_c^{k+1}.
\ee
Exchanging an order of the summation, we finally obtain
\be
{\cal F}_{\chi=1}(S,\alpha_i)
=-\sum_{i=1}^{N_f}\sum_{k=1}^{\infty}
\frac{(2k-1)!}{(k+1)!k!}{\alpha_i}^k S^{k+1},\\
\label{series}
\ee
where $\alpha_i$ is the same one defined in the previous section.

We now consider the series in (\ref{series}) which is written by a generalized hypergeometric function and has a branch cut discontinuity running from $\frac{1}{4}$ to $\infty$ in the complex plane. This hypergeometric series could be represented by a standard function as
\be
\sum_{k=1}^{\infty}
\frac{(2k-1)!}{(k+1)!k!}x^k
=\frac{1}{2}-\frac{1}{2f^{(\pm)}(x)}-\log\,f^{(\pm)}(x),
\label{series2}
\ee
where $f^{(\pm)}(x)$ are two solutions of the quadratic equation $f^2{-}f{+}x=0$, namely, defined by
\be
f^{(\pm)}(x)\equiv \frac{1\pm\sqrt{1-4x}}{2}.
\label{fpm}
\ee
The sign of $f^{(\pm)}(x)$ depends on which branch $x$ exists.

If we use the above functional representation of the series (\ref{series2}), we obtain the free energy from the connected diagrams with one boundary
\be
{\cal F}_{\chi=1}(S,\alpha_i)
=-S\sum_{i=1}^{N_f}\left(
\frac{1}{2}-\frac{1}{2f^{(\pm)}(\alpha_i S)}-\log\,f^{(\pm)}(\alpha_i S)
\right),
\label{F1(S)}
\ee
where the signs are suitably chosen for each $i$. We notice that the vacuum of \cite{ACFH,BR} represents only the case that all signs are the same, namely the $r{=}0$ case. This structure of vacua is preserved even though the mass matrix is proportional to the identity matrix.

Therefore finally we obtain the effective superpotential from the above matrix integration in terms of the glueball field
\bea
W_{\rm eff}(S) &=&
N_c \frac{\del {\cal F}_{\chi=2}(S)}{\del S}
-2\pi i \tau_0 S
+{\cal F}_{\chi=1}(S)\nn\\
&=&S\left[N_c(1{-}\log(S/\Lambda^3))
-\sum_{i=1}^{N_f}\left(
\frac{1}{2}-\frac{1}{2f^{(\pm)}(\alpha_i S)}-\log\,f^{(\pm)}(\alpha_i S)
\right)
\right].
\eea
The F-flatness equation with respect to $S$ gives
\be
S^{N_c}
=\frac{1}{2^{N_f}}\Lambda^{3N_c}
\prod_i \left(1+\sqrt{1-4\alpha_i S}\right)
\prod_{i'} \left(1-\sqrt{1-4\alpha_{i'} S}\right),
\label{S=prod}
\ee
where $i$ and $i'$ run for index of the positive and negative roots, respectively.

This equation is exactly equivalent to eq.~(\ref{X=prod}) identifying $X$ with $S$. So these theories have obviously the same structure of the vacuum solutions. If we especially consider the case that all $\alpha_i$'s are identical, the vacuum equation has $2N_c{-}N_f$ solutions with each fixed choice of the signs and the number of the independent way to choose the signs is $2^{N_f-1}$ for the same reason mentioned in the previous section. So we conclude that the Witten index of this theory is
\be
\Tr (-1)^F = (2N_c-N_f) 2^{N_f-1},
\ee
which agrees with (\ref{WittenM}).

We note here that the above vacuum structure still holds in the limit of $m_i\rightarrow 0$ and $\mu \rightarrow \infty$ with $\alpha_i$ fixed, namely we can take a massless limit of quarks without the adjoint scalar fields $\Phi$ in $N=1$ theory. However if take the limit of $\mu\rightarrow \infty$ first, which corresponds to $\alpha_i \rightarrow 0$, then the vacuum equation reduces to (no more choice of the signs)
\be
S^{N_c}=(\det m)\Lambda_1^{3N_c -N_f},
\ee
which gives only $N_c$ vacua as well as the pure $U(N_c)$ case. We must be careful that these limits are not commutable.

%
%
%
%

\section{Complete derivation of the ADS superpotential from matrix model}

In this section, we will see more explicit correspondence between the effective meson theory and the direct matrix model calculation with a complete proof. We again begin with the matrix integral
\be
Z=\frac{{\cal N}_{Q}}{\Vol(U(N_c))}
\int [d\Phi][dQ][d\Qt]
\exp\left\{
-\frac{1}{g_s}W_{\rm tree}(\Phi,Q,\Qt)
\right\},
\label{ele mat}
\ee
where ${\cal N}_Q\equiv\left(\det (m/2\pi g_s)\right)^{N_c}$ is a normalization constant which is determined by the gaussian integral over $Q$ and $\Qt$.

To compare with the effective theory written by the mesonic variables $M_i^j=Q_i^a\Qt^j_a$, we consider a change of variables in the superpotential by insertion of a matrix valued delta function
\be
e^{-\frac{1}{g_s}W_{\rm tree}(\Phi,Q,\Qt)}
=\int [dM]\delta(M-Q\Qt)
e^{-\frac{1}{g_s}W_{\rm tree}(\tilde{\Phi},M)},
\label{delta}
\ee
where
\be
W_{\rm tree}(\tilde{\Phi},M)
= \mu\Tr\tilde{\Phi}^2
- \frac{1}{2\mu}\Tr M^2 + \Tr(mM),
\ee
with $
\tilde{\Phi}_a^b\equiv
\Phi_a^b
+\frac{1}{\sqrt{2}\mu}\Qt^b_i Q_a^i
$.
The matrix valued delta function is defined by
\be
\delta(M-Q\Qt)
=\frac{1}{(2\pi)^{N_f^2}}
\int [dT] \exp\left\{i\Tr\,T(M-Q\Qt)\right\},
\ee
where $T$ is an $N_f \times N_f$ matrix. Note that this variable change is valid for $N_c>N_f$ since each matrices have the same degrees of freedom. If $N_c\leq N_f$, then we have to introduce more constraints of variables like baryons.

Substituting (\ref{delta}) into (\ref{ele mat}) and integrating gaussian integrals of $Q$, $\Qt$ and $\Phi$, we find
\bea
Z&=&\frac{{\cal N}_{Q}}{(2\pi)^{N_f^2}\Vol(U(N_c))}
\int [d\tilde{\Phi}][dQ][d\Qt][dM][dT]
\exp\left\{
-\frac{1}{g_s}W_{\rm tree}(\tilde{\Phi},M)
+i\Tr\,T(M-Q\Qt)
\right\}\nn\\
&=&\frac{{\cal N}_{Q}\left(\frac{2\pi g_s}{\mu}\right)^{\frac{1}{2}N_c^2}}{(2\pi)^{N_f^2-N_cN_f}\Vol(U(N_c))}
\int [dM][dT]
\left(\det(iT)\right)^{-N_c}\exp\left\{
-\frac{1}{g_s}
W_{\rm tree}(M)
+i\Tr\,TM
\right\},\nn\\
\label{MM+TM}
\eea
where $W_{\rm tree}(M)=\Tr\left(-\frac{1}{2\mu}M^2+mM\right)$ is the tree superpotential of the mesonic part only.

Let us consider the following integration over $T$ first. We define an integral
\be
I_{N_c,N_f}(M)\equiv
\int [dT]
\left(\det(iT)\right)^{-N_c}
e^{i\Tr\,TM}.
\ee
To perform this integral, we decompose the matrices $T$ and $M$ into smaller parts
\be
T=\left(
\ba{cc}
\tau & \tv^\dag \\
\tv & T'
\ea
\right),
\quad
M=\left(
\ba{cc}
\rho & \vv^\dag \\
\vv & M'
\ea
\right),
\ee
where $\tau$ and $\rho$ are scalars, $\tv$ and $\vv$ are $N_f-1$ vectors and $T'$ and $M'$ are $(N_f{-}1)\times(N_f{-}1)$ matrices. This decomposition leads to the decomposition of the determinant of $iT$
\be
\det(iT)=
\left(
i\tau+\tv^\dag (iT')^{-1} \tv
\right)
\det(iT').
\ee

Now we can integrate over the real scalar $\tau$ by a residue integral
\bea
I_{N_c,N_f}(M)
&=&\frac{2\pi}{\Gamma(N_c)}
\rho^{N_c-1}
\int [dT'] \left(\det(iT')\right)^{-N_c}
\exp\left\{
i\Tr\,T'M'\right\}\nn\\
&&\qquad\qquad\times
\int d\tv d\tv^\dag
\exp\left\{
-\rho(\tv^\dag+\av^\dag)(iT')^{-1}(\tv+\av)
-\frac{1}{\rho}\vv^\dag(iT')\vv
\right\},
\eea
where $\av=-\frac{i}{\rho}(iT')\vv$. Next if we notice that the integration over $\tv$ and $\tv^\dag$ is a gaussian integral with a constant shift, we obtain the following recursive relation
\bea
I_{N_c,N_f}(M)&=&\frac{(2\pi)^{\frac{1}{2}(N_f+1)}}{\Gamma(N_c)}\rho^{N_c-N_f}
\int [dT']\left(\det(iT')\right)^{N_c-1}
\exp\left\{
i\Tr\,T'\left(M'-\frac{1}{\rho}\vv\vv^\dag\right)
\right\}\nn\\
&=&\frac{(2\pi)^{\frac{1}{2}(N_f+1)}}{\Gamma(N_c)}\rho^{N_c-N_f}
I_{N_c-1,N_f-1}\left(M'-\frac{1}{\rho}\vv\vv^\dag\right).
\eea

We can easily solve the above recursive relation from the initial condition $I_{N_c-N_f+1,1}(\rho)=\frac{2\pi}{\Gamma(N_c-N_f+1)}\rho^{N_c-N_f}$ if and only if $N_c>N_f$. So we find
\bea
I_{N_c,N_f}(M)&=&
\frac{(2\pi)^{\frac{1}{2}N_f^2+\frac{1}{2}N_f}}{\prod_{j=1}^{N_f}\Gamma(N_c-j+1)}
\left(\det\, M\right)^{N_c-N_f}\nn\\
&=&
(2\pi)^{\frac{1}{2}N_f^2+\frac{1}{2}N_f}
\frac{G_2(N_c-N_f+1)}{G_2(N_c+1)}
\left(\det\, M\right)^{N_c-N_f},
\label{INN}
\eea
where $G_2(z)$ is the Barnes function defined by
\be
G_2(z+1) = \Gamma(z)G_2(z),\quad G_2(1)=1,
\ee
and which has an asymptotic expansion at the large $z$
\be
\log G_2(z+1)
\simeq \frac{z^2}{2}\left(\log z-\frac{3}{2}\right)
+\frac{z}{2}\log 2\pi
-\frac{1}{12}\log z.
\ee

We would like to emphasize here that the formula (\ref{INN}) does not contain a term $e^{-\Tr M}$ in contrast with the discussion in \cite{DJ,Fyodorov,JN}. This term has a difficulty in a field theoretical sense since it is proportional to a mass term of the meson field and shifts the meson mass. However, there is no problem in our arguments.

Now if we recall the volume of the $U(N)$ gauge group is given by \cite{Macdonald,OV}
\be
\Vol(U(N))=\frac{(2\pi)^{\frac{1}{2}N^2+\frac{1}{2}N}}{G_2(N+1)},
\ee
plugging back the formula (\ref{INN}) into eq.~(\ref{MM+TM}), we obtain
\be
Z=\frac{{\cal N}_{Q}\left(\frac{2\pi g_s}{\mu}\right)^{\frac{1}{2}N_c^2}}
{\Vol(U(N_c{-}N_f))}
\int [dM]
\left(\det\,M\right)^{N_c-N_f}
e^{-\frac{1}{g_s}W_{\rm tree}(M)}.
\label{intM}
\ee
The volume factor is changed from $U(N_c)$ to $U(N_c-N_f)$. This means that the Veneziano-Yankielowicz term is modified to $(N_c-N_f)S\log S$ due to the matter contribution.

We first consider to complete the above integration over $M$. In the large $N_c$, $N_f$ limit with $N_f/N_c$ fixed, the above integration over $M$ can be evaluated by the saddle point approximation. If we again diagonalize $M$ like (\ref{diagM})\footnote{This does not bring about the Vandermonde determinant since $M$ is always diagonalized by the fixed $U(N_f)$ global symmetry.}, the integral of $M$ is dominated at the solution of the saddle point equations
\be
-\frac{1}{\mu}\lambda_i+m_i-S \lambda_i^{-1}=0.
\ee
This quadratic equations has two solutions for each $\lambda_i$
\be
\lambda_i^{(\pm)}=\mu\,m_i
\left(
\frac{1\pm\sqrt{1-4\alpha_i S}}{2}
\right).
\ee
Picking terms proportional to $1/g_s$ from the approximated $M$ integral part and the volume factor in $-\log\,Z$, we find exactly
\be
{\cal F}_{\chi=1}(S,\alpha_i)
=-S \sum_{i=1}^{N_f}\left(
\frac{1}{2}
-\frac{1}{2f^{(\pm)}(\alpha_i S)}
-\log\,f^{(\pm)}(\alpha_i S)
\right),
\ee
where $f^{(\pm)}(x)$ are defined in (\ref{fpm}) and signs are properly chosen for each $i$.
This final result, of course, is the same as (\ref{F1(S)}) in the previous section since we have just performed the matrix integration over all variables in the alternative way.

We have here integrated all matrix variables to the end, so we have skipped  the intermediate mesonic effective theory with the ADS superpotential. To obtain the ADS superpotential from this matrix integral, we have to forget temporally the integral over $M$ in (\ref{intM}) and use the DV procedure before the integral of $M$. We can justify this procedure since the size of $M$ depends only on $N_f$ and the $M$ integral does not contribute to the function of $S$. We find the expansion of the free energy from $-\log Z$ and get
\bea
W_{\rm eff}(S,M) &=&
N_c \frac{\del {\cal F}_{\chi=2}(S)}{\del S}
-2\pi i \tau_0 S
+{\cal F}_{\chi=1}(S)\nn\\
&=&S\left[N_c(1{-}\log(S/\Lambda^3))
-N_f(1{-}\log S)
-\log\left(\det(mM)\right)
\right]
+W_{\rm tree}(M).\nn\\
\label{W(S,M)}
\eea
Next we consider the minimaizing of $S$, namely if we solve the F-flatness condition with respect to $S$ and substitute into the solution to (\ref{W(S,M)}), we obtain the effective meson theory
\be
W_{\rm eff}(M)=W_{\rm tree}(M) + W_{\rm ADS}(M),
\label{eff meson}
\ee
where $W_{\rm ADS}$ is exactly the same as (\ref{ADS}) with the identification of the scale $\Lambda_1^{3N_c-N_f} = \Lambda^{3N_c}/\det m$. If we minimize the superpotential (\ref{eff meson}), we apparently see the same vacuum structure as the case where all matrix variables are integrated since we just minimize with respect to $M$ and $S$ in each case.

Now we consider the special case of $N_c{=}N_f$. We are able to use the above argument for $N_c{=}N_f$ excluding the baryons, $B=\det Q$ and $\tilde{B}=\det \tilde{Q}$. All logarithmic functions of $S$ disappear from eq.~(\ref{W(S,M)}) in this case. Then we have
\be
W_{\rm eff}(S,M) = S
\log\left(
\frac{\Lambda_1^{2 N_c}}{\det M}
\right)
+W_{\rm tree}.
\ee
From the F-flatness equation of $S$, we obtain the following equation without any gluino condensation
\be
\det M = \Lambda_1^{2N_c}.
\ee
Therefore we obtain a quantum modification of the classical moduli space from the matrix model \cite{Berenstein} when $B=0$ or $\tilde{B}=0$.

Finally we note that the integration of the delta function exchanges the volume factor of $U(N_c)$ for one of $U(N_c{-}N_f)$. If we naively apply the formula (\ref{INN}) to the case of $N_c < N_f$, we can rewrite eq.~(\ref{intM}) up to some constants as
\be
Z \sim \frac{1}
{\Vol(U(N_f{-}N_c))}
\int [dM][dq][d\tilde{q}]
e^{-\frac{1}{g_s}\left(W_{\rm tree}(M)+\tilde{q} M q\right)},
\ee
where $q$ and $\tilde{q}$ are $(N_f{-}N_c) \times N_f$ matrices, namely, which represent the fundamental flavors of $U(N_f{-}N_c)$ gauge theory.
This fact strongly suggests the appearance of the Seiberg duality from the matrix models \cite{Feng,FH}. Moreover, we comment that the above rough derivation of Seiberg duality is valid even without any mass deformation. However, the degrees of freedom of the quark matrices $Q$ and $\Qt$ can not be replaced by only the mesonic variables for $N_c \leq N_f$. So we should need more careful analysis with some constraints or baryon matrices \cite{BRT}.

\section*{Acknowledgements}

I would like to thank N.~Dorey, T.~J.~Hollowood, S.~P.~Kumar, P.~de Madeiros, H.~Suzuki and T~.Yokono for useful discussions and comments. This work is supported in part by PPARC for Research into Applications of Quantum Theory to Problems in Fundamental Physics.

\input refs.tex

\end{document}

%% file: fig.tex
\unitlength 0.1in
\begin{picture}( 38.2300, 16.1800)(  9.9000,-25.3800)
%
\special{pn 13}%
\special{sh 0.300}%
\special{ar 1800 1730 810 810  0.0000000 6.2831853}%
%
\special{pn 13}%
\special{sh 0}%
\special{ar 2146 1334 178 178  0.0000000 6.2831853}%
%
\special{pn 13}%
\special{sh 0}%
\special{ar 1536 1292 178 178  0.0000000 6.2831853}%
%
\special{pn 13}%
\special{sh 0}%
\special{ar 1312 1816 178 178  0.0000000 6.2831853}%
%
\special{pn 13}%
\special{sh 0}%
\special{ar 1746 2222 178 178  0.0000000 6.2831853}%
%
\special{pn 13}%
\special{sh 0}%
\special{ar 2264 1900 178 178  0.0000000 6.2831853}%
%
\special{pn 13}%
\special{sh 0}%
\special{ar 1816 1740 178 178  0.0000000 6.2831853}%
%
\special{pn 8}%
\special{pa 1508 1088}%
\special{pa 1564 1152}%
\special{fp}%
%
\special{pn 13}%
\special{pa 1508 1144}%
\special{pa 1564 1088}%
\special{fp}%
%
\special{pn 8}%
\special{pa 1516 1438}%
\special{pa 1572 1502}%
\special{fp}%
%
\special{pn 13}%
\special{pa 1516 1494}%
\special{pa 1572 1438}%
\special{fp}%
%
\special{pn 8}%
\special{pa 2118 1138}%
\special{pa 2174 1200}%
\special{fp}%
%
\special{pn 13}%
\special{pa 2118 1194}%
\special{pa 2174 1138}%
\special{fp}%
%
\special{pn 8}%
\special{pa 2118 1474}%
\special{pa 2174 1536}%
\special{fp}%
%
\special{pn 13}%
\special{pa 2118 1530}%
\special{pa 2174 1474}%
\special{fp}%
%
\special{pn 8}%
\special{pa 1788 1544}%
\special{pa 1844 1606}%
\special{fp}%
%
\special{pn 13}%
\special{pa 1788 1600}%
\special{pa 1844 1544}%
\special{fp}%
%
\special{pn 8}%
\special{pa 1782 1886}%
\special{pa 1838 1950}%
\special{fp}%
%
\special{pn 13}%
\special{pa 1782 1942}%
\special{pa 1838 1886}%
\special{fp}%
%
\special{pn 8}%
\special{pa 2244 1704}%
\special{pa 2300 1768}%
\special{fp}%
%
\special{pn 13}%
\special{pa 2244 1760}%
\special{pa 2300 1704}%
\special{fp}%
%
\special{pn 8}%
\special{pa 2244 2040}%
\special{pa 2300 2104}%
\special{fp}%
%
\special{pn 13}%
\special{pa 2244 2096}%
\special{pa 2300 2040}%
\special{fp}%
%
\special{pn 8}%
\special{pa 1270 1620}%
\special{pa 1326 1684}%
\special{fp}%
%
\special{pn 13}%
\special{pa 1270 1676}%
\special{pa 1326 1620}%
\special{fp}%
%
\special{pn 8}%
\special{pa 1284 1956}%
\special{pa 1340 2020}%
\special{fp}%
%
\special{pn 13}%
\special{pa 1284 2012}%
\special{pa 1340 1956}%
\special{fp}%
%
\special{pn 8}%
\special{pa 1718 2026}%
\special{pa 1774 2090}%
\special{fp}%
%
\special{pn 13}%
\special{pa 1718 2082}%
\special{pa 1774 2026}%
\special{fp}%
%
\special{pn 8}%
\special{pa 1726 2370}%
\special{pa 1782 2432}%
\special{fp}%
%
\special{pn 13}%
\special{pa 1726 2426}%
\special{pa 1782 2370}%
\special{fp}%
\put(15.3600,-10.3200){\makebox(0,0){$\lambda$}}%
\put(15.4300,-15.2900){\makebox(0,0){$\lambda$}}%
\put(18.0900,-26.7700){\makebox(0,0){$N_c h S^{h-1}$}}%
%
\special{pn 13}%
\special{sh 0.300}%
\special{ia 4004 1730 810 810  0.0000000 6.2831853}%
\special{ar 4004 1730 810 810  0.0000000 0.0741656}%
\special{ar 4004 1730 810 810  0.1186650 0.1928307}%
\special{ar 4004 1730 810 810  0.2373300 0.3114957}%
\special{ar 4004 1730 810 810  0.3559951 0.4301607}%
\special{ar 4004 1730 810 810  0.4746601 0.5488257}%
\special{ar 4004 1730 810 810  0.5933251 0.6674907}%
\special{ar 4004 1730 810 810  0.7119901 0.7861557}%
\special{ar 4004 1730 810 810  0.8306551 0.9048208}%
\special{ar 4004 1730 810 810  0.9493201 1.0234858}%
\special{ar 4004 1730 810 810  1.0679852 1.1421508}%
\special{ar 4004 1730 810 810  1.1866502 1.2608158}%
\special{ar 4004 1730 810 810  1.3053152 1.3794808}%
\special{ar 4004 1730 810 810  1.4239802 1.4981459}%
\special{ar 4004 1730 810 810  1.5426452 1.6168109}%
\special{ar 4004 1730 810 810  1.6613103 1.7354759}%
\special{ar 4004 1730 810 810  1.7799753 1.8541409}%
\special{ar 4004 1730 810 810  1.8986403 1.9728059}%
\special{ar 4004 1730 810 810  2.0173053 2.0914710}%
\special{ar 4004 1730 810 810  2.1359703 2.2101360}%
\special{ar 4004 1730 810 810  2.2546354 2.3288010}%
\special{ar 4004 1730 810 810  2.3733004 2.4474660}%
\special{ar 4004 1730 810 810  2.4919654 2.5661310}%
\special{ar 4004 1730 810 810  2.6106304 2.6847960}%
\special{ar 4004 1730 810 810  2.7292954 2.8034611}%
\special{ar 4004 1730 810 810  2.8479604 2.9221261}%
\special{ar 4004 1730 810 810  2.9666255 3.0407911}%
\special{ar 4004 1730 810 810  3.0852905 3.1594561}%
\special{ar 4004 1730 810 810  3.2039555 3.2781211}%
\special{ar 4004 1730 810 810  3.3226205 3.3967862}%
\special{ar 4004 1730 810 810  3.4412855 3.5154512}%
\special{ar 4004 1730 810 810  3.5599506 3.6341162}%
\special{ar 4004 1730 810 810  3.6786156 3.7527812}%
\special{ar 4004 1730 810 810  3.7972806 3.8714462}%
\special{ar 4004 1730 810 810  3.9159456 3.9901112}%
\special{ar 4004 1730 810 810  4.0346106 4.1087763}%
\special{ar 4004 1730 810 810  4.1532756 4.2274413}%
\special{ar 4004 1730 810 810  4.2719407 4.3461063}%
\special{ar 4004 1730 810 810  4.3906057 4.4647713}%
\special{ar 4004 1730 810 810  4.5092707 4.5834363}%
\special{ar 4004 1730 810 810  4.6279357 4.7021014}%
\special{ar 4004 1730 810 810  4.7466007 4.8207664}%
\special{ar 4004 1730 810 810  4.8652658 4.9394314}%
\special{ar 4004 1730 810 810  4.9839308 5.0580964}%
\special{ar 4004 1730 810 810  5.1025958 5.1767614}%
\special{ar 4004 1730 810 810  5.2212608 5.2954265}%
\special{ar 4004 1730 810 810  5.3399258 5.4140915}%
\special{ar 4004 1730 810 810  5.4585909 5.5327565}%
\special{ar 4004 1730 810 810  5.5772559 5.6514215}%
\special{ar 4004 1730 810 810  5.6959209 5.7700865}%
\special{ar 4004 1730 810 810  5.8145859 5.8887515}%
\special{ar 4004 1730 810 810  5.9332509 6.0074166}%
\special{ar 4004 1730 810 810  6.0519159 6.1260816}%
\special{ar 4004 1730 810 810  6.1705810 6.2447466}%
%
\special{pn 13}%
\special{sh 0}%
\special{ar 4350 1334 178 178  0.0000000 6.2831853}%
%
\special{pn 13}%
\special{sh 0}%
\special{ar 3742 1292 178 178  0.0000000 6.2831853}%
%
\special{pn 13}%
\special{sh 0}%
\special{ar 3518 1816 178 178  0.0000000 6.2831853}%
%
\special{pn 13}%
\special{sh 0}%
\special{ar 3952 2222 178 178  0.0000000 6.2831853}%
%
\special{pn 13}%
\special{sh 0}%
\special{ar 4470 1900 178 178  0.0000000 6.2831853}%
%
\special{pn 13}%
\special{sh 0}%
\special{ar 4022 1740 178 178  0.0000000 6.2831853}%
%
\special{pn 8}%
\special{pa 3714 1088}%
\special{pa 3770 1152}%
\special{fp}%
%
\special{pn 13}%
\special{pa 3714 1144}%
\special{pa 3770 1088}%
\special{fp}%
%
\special{pn 8}%
\special{pa 3720 1438}%
\special{pa 3776 1502}%
\special{fp}%
%
\special{pn 13}%
\special{pa 3720 1494}%
\special{pa 3776 1438}%
\special{fp}%
%
\special{pn 8}%
\special{pa 4322 1138}%
\special{pa 4378 1200}%
\special{fp}%
%
\special{pn 13}%
\special{pa 4322 1194}%
\special{pa 4378 1138}%
\special{fp}%
%
\special{pn 8}%
\special{pa 4322 1474}%
\special{pa 4378 1536}%
\special{fp}%
%
\special{pn 13}%
\special{pa 4322 1530}%
\special{pa 4378 1474}%
\special{fp}%
%
\special{pn 8}%
\special{pa 3994 1544}%
\special{pa 4050 1606}%
\special{fp}%
%
\special{pn 13}%
\special{pa 3994 1600}%
\special{pa 4050 1544}%
\special{fp}%
%
\special{pn 8}%
\special{pa 3986 1886}%
\special{pa 4042 1950}%
\special{fp}%
%
\special{pn 13}%
\special{pa 3986 1942}%
\special{pa 4042 1886}%
\special{fp}%
%
\special{pn 8}%
\special{pa 4448 1704}%
\special{pa 4504 1768}%
\special{fp}%
%
\special{pn 13}%
\special{pa 4448 1760}%
\special{pa 4504 1704}%
\special{fp}%
%
\special{pn 8}%
\special{pa 4448 2040}%
\special{pa 4504 2104}%
\special{fp}%
%
\special{pn 13}%
\special{pa 4448 2096}%
\special{pa 4504 2040}%
\special{fp}%
%
\special{pn 8}%
\special{pa 3476 1620}%
\special{pa 3532 1684}%
\special{fp}%
%
\special{pn 13}%
\special{pa 3476 1676}%
\special{pa 3532 1620}%
\special{fp}%
%
\special{pn 8}%
\special{pa 3490 1956}%
\special{pa 3546 2020}%
\special{fp}%
%
\special{pn 13}%
\special{pa 3490 2012}%
\special{pa 3546 1956}%
\special{fp}%
%
\special{pn 8}%
\special{pa 3924 2026}%
\special{pa 3980 2090}%
\special{fp}%
%
\special{pn 13}%
\special{pa 3924 2082}%
\special{pa 3980 2026}%
\special{fp}%
%
\special{pn 8}%
\special{pa 3930 2370}%
\special{pa 3986 2432}%
\special{fp}%
%
\special{pn 13}%
\special{pa 3930 2426}%
\special{pa 3986 2370}%
\special{fp}%
\put(37.4100,-10.3200){\makebox(0,0){$\lambda$}}%
\put(37.4800,-15.2900){\makebox(0,0){$\lambda$}}%
\put(40.0700,-26.7000){\makebox(0,0){$N_f S^{h'}$}}%
\end{picture}%

%% file: refs.tex
%
%
\newpage


\newcommand{\mathph}[1]{\href{http://xxx.lanl.gov/abs/math-ph/#1}{\tt math-ph/#1}}

\newcommand{\lit}[3]{#1, ``{\it #2}'', #3.}
\newcommand{\lits}[2]{#1, #2.}